%% file: ICIP_2020.tex
\documentclass{article}
\usepackage{spconf,amsmath,graphicx}

\usepackage{cite}
\usepackage{amsmath,amssymb,amsfonts}
\usepackage{algorithmic}
\usepackage{textcomp}
\usepackage{xcolor}
\usepackage{booktabs}
\usepackage{graphicx}
\usepackage{caption}
\usepackage{subcaption}
\usepackage{siunitx, soul}
\usepackage{adjustbox}
\usepackage{url} % hyperref works too
\usepackage{multirow}

\usepackage[acronym,toc,nomain]{glossaries}
\input{Glossary.tex}
\let\oldtabular=\tabular
\def\tabular{\small\oldtabular}

\newcommand{\Figure}[1] {\textbf{Fig.~#1}}
\newcommand{\Table}[1] {\textbf{Table~#1}}

\title{Quality-driven Dynamic VVC Frame Partitioning for efficient Parallel Processing}
\name{Thomas AMESTOY$^{\star,\dagger}$, Wassim HAMIDOUCHE$^\star$, Cyril BERGERON$^\dagger$ and Daniel MENARD$^\star$
	\thanks{This project has received funding from Bpifrance Financement under grant DOS0061463/00 (EFIGI FUI project).}
}
\address{$^\star$ Univ Rennes, INSA Rennes, CNRS, IETR - UMR 6164, Rennes, France. Emails: firstname.lastname@insa-rennes.fr \\
	$^\dagger$ Thales SIX GTS France, HTE/STR/MMP	Gennevilliers, France. Emails: firstname.lastname@thalesgroup.com 
}

\begin{document}
\ninept
\setlength{\textfloatsep}{3pt}% Remove \textfloatsep
\setlength{\intextsep}{3pt}
\maketitle

\begin{sloppypar}
\begin{abstract}
	\acrshort{vvc} is the next generation video coding standard, offering coding capability beyond \acrshort{hevc} standard. 
	The high computational complexity of the latest video coding standards requires high-level parallelism techniques, in order to achieve real-time and low latency encoding and decoding.
%	In \acrshort{hevc} for instance, the tile grid partitioning of a frame allows simultaneous processing of rectangular regions of the frame with independent threads.
%	In \acrshort{vvc}, further partition of tile grid into a horizontal sub-grid of \glspl{rs} is enabled increasing the partitioning flexibility. 
%	This paper investigates dynamic \gls{trs} partitioning solution that benefits from this increased flexibility.
	\acrshort{hevc} and \acrshort{vvc} include tile grid partitioning that allows to process simultaneously rectangular regions of a frame with independent threads.
	The tile grid may be further partitioned into a horizontal sub-grid of \glspl{rs}, increasing the partitioning flexibility. 
	The dynamic \gls{trs} partitioning solution proposed in this paper benefits from this flexibility.
%	In this paper, a dynamic \gls{trs} partitioning solution that benefits from this increased flexibility is proposed for \acrshort{vvc}.
	The \gls{trs} partitioning is carried-out at the frame level, taking into account both spatial texture of the content and encoding times of previously encoded frames.
	The proposed solution searches the best partitioning configuration that minimizes the trade-off between multi-thread encoding time and encoding quality loss. 
	Experiments prove that the proposed solution, compared to uniform \gls{trs} partitioning, significantly decreases multi-thread encoding time, with slightly better encoding quality.
\end{abstract}
\glsresetall

\section{Introduction}\label{section:intro}
In recent years, the democratization of multimedia applications, coupled with the emergence of high resolution and new video formats (8K, 360\textdegree), has led to a drastic increase in the volume of exchanged video content~\cite{cisco_global_2021_forecast_highlights_2016}.
This increasing need for higher compression rates prompted the \gls{jvet} 
%to investigate several new coding solutions, in order 
to develop a new video coding standard called \gls{vvc} with coding capability beyond \gls{hevc}~\cite{sullivan_overview_2012}. The bit-rate savings brought by \gls{vvc}\cite{sidaty_compression_2019} are however coupled with a considerable encoding computational complexity increase.
This latter is estimated to 10 and 27 times \gls{hevc} computational complexity in Inter and Intra coding configuration, respectively~\cite{bossen_jvet-p0003_2019}.
In real-time implementations of \gls{vvc} codec, intense parallel processing will therefore be mandatory to achieve real-time encoding and decoding.

\begin{figure}[ht]
	\centering
	\begin{minipage}[b]{.49\linewidth}
		\begin{subfigure}[b]{\linewidth}
			\includegraphics[width=0.98\linewidth]{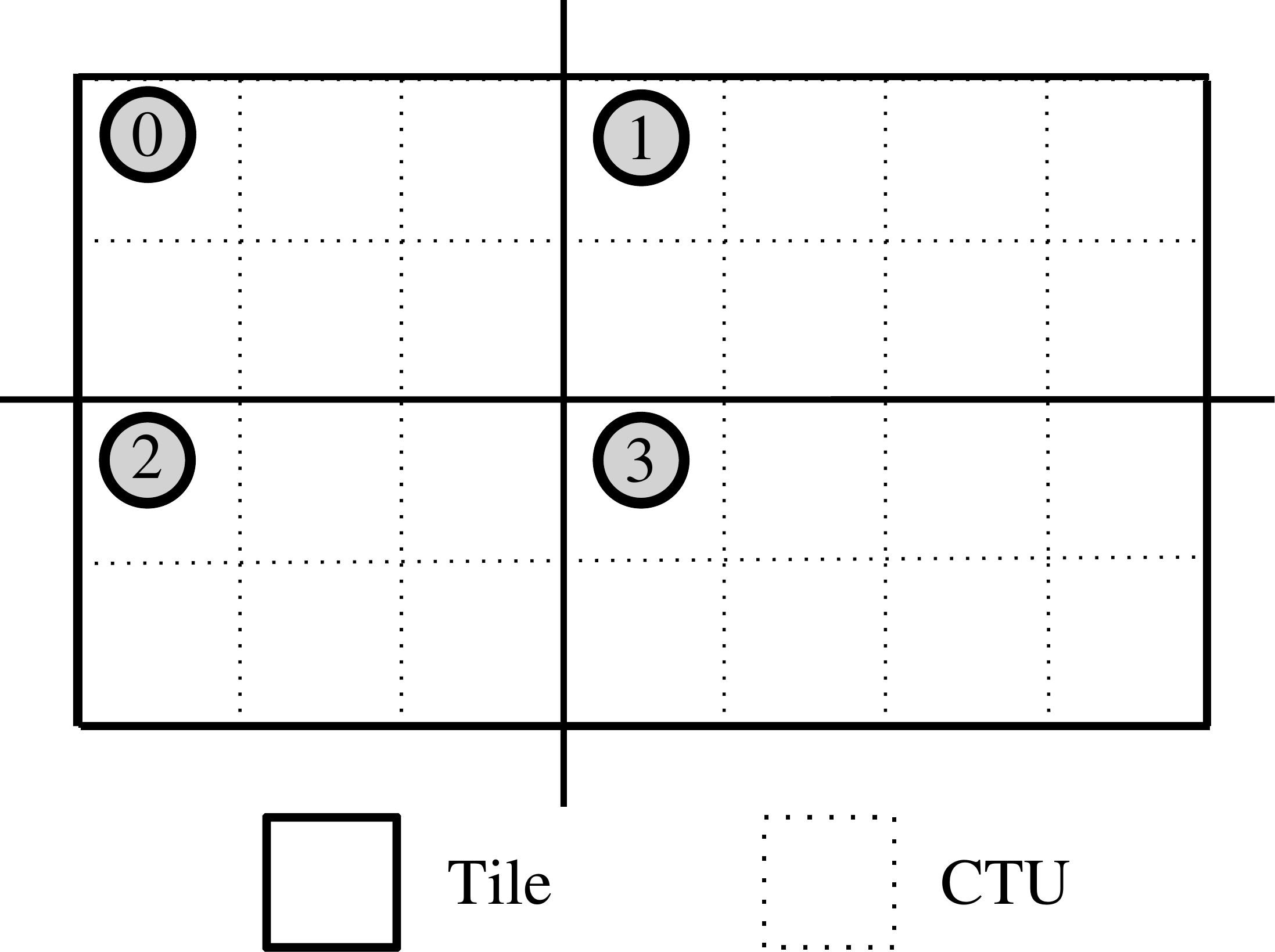}
			\caption{Grid of 4 tiles.\\Tiles labeled  from 0 to 3 }
			\label{fig:hevc_tile}
		\end{subfigure}
	\end{minipage} \hfill
	\begin{minipage}[b]{.49\linewidth}
		\begin{subfigure}[b]{\linewidth}
			\includegraphics[width=0.98\linewidth]{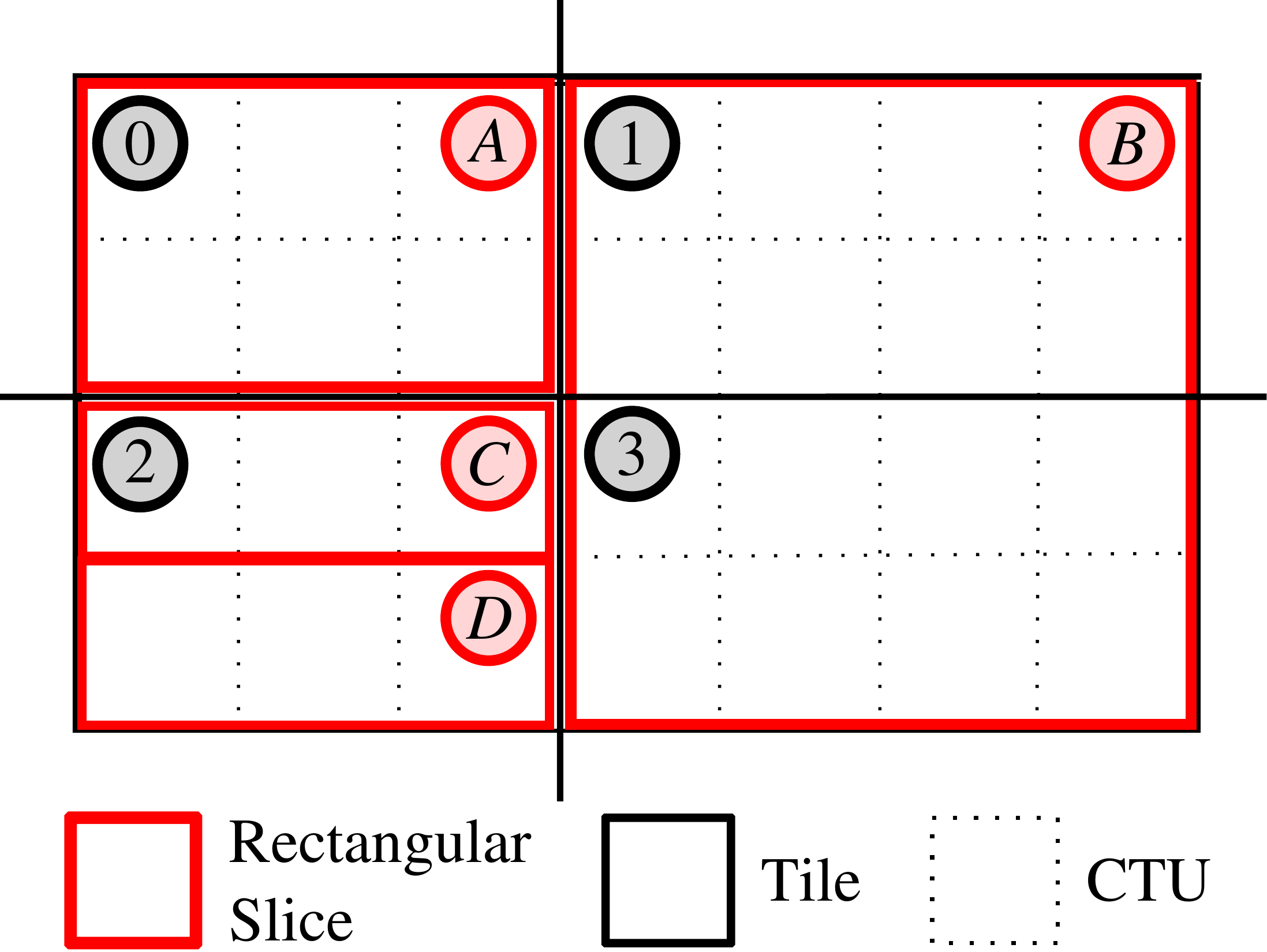}
			\caption{Tiles combined with 4 \glspl{rs}. \glspl{rs} labeled  from A to D.}
			\label{fig:slice_tile}
		\end{subfigure}
	\end{minipage}
	\caption{Illustration of tile partitioning in \gls{hevc} and \gls{trs} partitioning in \gls{vvc}.}
	\label{fig:tiles_hevc_vs_vvc}
\end{figure}

Techniques of video parallel processing essentially operate at three levels of parallelism: data level, frame level and high-level.
The data level parallelism techniques are applied on elementary operations, and no encoding quality is lost compared to sequential encoding. They include among other techniques relying on \gls{simd} architectures~\cite{bross_hevc_2013}. Frame level and high-level parallelism operate at thread level.%; speed-up is achieved at the cost of encoding quality loss. 
The frame level techniques encode a group of frames in parallel where each thread is assigned to a single frame~\cite{hamidouche_4k_2016}.
The encoding time of a single frame is not reduced with frame level techniques, i.e. the latency is not reduced.
In high-level parallelism techniques, the threads operate on continuous regions of the frame, as tiles or slices~\cite{misra_overview_2013}.
Tiles and slices are independently encodable and decodable, allowing several threads to process simultaneously the same frame.  
%The encoding and decoding times of every frame is therefore reduced, improving equally both speed-up and latency. 
These techniques improve equally both speed-up and latency. 
However, by enabling independent processing of frame regions, prediction dependencies across boundaries are broken and entropy encoding state is reinitialized for each region.
These restrictions lead to an encoding quality loss compared to the encoding of the non-partitioned sequence.
The encoding quality decreases with the number of independent regions of the frame, as has been measured in \gls{hevc} by \textit{Chin et al.}~\cite{chi_ching_chi_parallel_2012}.
In \gls{hevc} and \gls{vvc} standards, only grid shaped tile partitioning is allowed, as shown by \Figure{\ref{fig:hevc_tile}}.
The tiles are delimited by the continuous black lines and the dashed lines correspond to the \gls{ctu} delimitation. %(blocks of size 128x128 typically for VVC \hl{y a pas 128 en HEVC}).
The tile partitioning forms a 2x2 grid and tiles are labelled from 0 to 3.
In order to increase the partitioning opportunities, \gls{vvc} combines the tile partitioning with the new concept of \glspl{rs}.
The partitioning combining tiles and \glspl{rs} is further called \gls{trs} partitioning.
\Figure{\ref{fig:slice_tile}} shows the \gls{trs} partitioning of a frame into the same 2x2 tile grid than \Figure{\ref{fig:hevc_tile}}, combined with 4 \glspl{rs}. The \glspl{rs} are delimited by the continuous red lines and are labeled from A to D. 
% and the dashed lines correspond to the  \gls{ctu} delimitations (blocks of 128x128 Luma samples typically).  
%and the 4 dashed segments correspond to the extensions of the tile grid lines, in order to facilitate the tile grid visualization.  
The \gls{rs} may contain one or several complete tiles, forming together a rectangular region of the frame. 
Moreover, as shown in the examples $C$ and $D$, a \gls{rs} may be a rectangular sub-region of the tile, composed of a number of complete and consecutive \gls{ctu} rows of a tile.
In this latter case, the \glspl{rs} allow to further partition the tile grid into a horizontal sub-grid, improving greatly the tile grid partitioning flexibility.
%As shown in \Figure{\ref{fig:slice_tile}}, a \gls{rs} in \gls{vvc} may contain a number of complete tiles, forming together a rectangular region of the frame (see \gls{rs} 1). 
%Moreover, as shown by the examples 0, 2 and 3, a \gls{rs} may also be composed of a number of complete and consecutive \gls{ctu} rows of a tile, which form collectively a rectangular region of the tile.
%In this latter case, the new concept of \gls{rs} allows to further partition the tile grid into an horizontal sub-grid, increasing greatly the \gls{trs} partitioning flexibility compared to tile grid partitioning in \gls{hevc}.

% The partition of a frame into independently encodable regions raises two distinct optimization issues: on one side the speed-up maximization (the speed-up being the ratio between multi-thread encoding time and sequential encoding time), on the other side the minimization of encoding quality losses.
The partitioning of a frame into tiles and \glspl{rs} raises two distinct optimization issues: on one side the multi-thread encoding time minimization (or speedup maximization), on the other side the minimization of encoding quality loss caused by the partitioning.
%By enabling independent processing of frame regions, coding dependencies across regions boundaries are broken and entropy encoding state is reinitialized for each region.
%These restrictions lead to encoding quality loss compared to the encoding of a non-partitioned frame.
In the literature, both issues have been addressed for \gls{hevc} tile partitioning.
%In order to minimize multi-thread encoding time, \textit{Storch et al}~\cite{storch_speedup-aware_2016} use the history encoding times of previous frames to determine the vertical and horizontal boundaries of current tile partitioning.
The multi-thread encoding time minimization is investigated by
\textit{Storch et al}~\cite{storch_speedup-aware_2016} and \textit{Koziri et al.}~\cite{koziri_heuristics_2017}.
They observe that the encoding time does not vary significantly from a \gls{ctu} to the co-located \gls{ctu} in the closest temporal frame.
Considering this temporal stability, the authors use the  encoding times of previous frames to determine the tile partitioning that minimizes the multi-thread encoding time.
In~\cite{ahn_complexity_2013}, the time estimator for each \gls{ctu} is computed based on previously encoded frame \gls{ctu} statistics (number of Skip, Inter, Intra blocks for instance).
%This time estimator is also used to minimize multi-thread encoding time.
Authors in~\cite{blumenberg_adaptive_2013,jin_clustering-based_2016} minimize the encoding quality loss induced by the tile partitioning by analyzing the \gls{ctu} luminance variances of the frame.
The technique proposed in~\cite{malossi_adjusting_2016} focuses on the particular case of variable number of available cores. The encoding loss is lowered in some cases by setting a number of tiles inferior to the number of available cores.
%However, the related works on \gls{hevc} tile partitioning only address encoding time minimization and encoding quality maximization independently, without trying to combine them as a trade-off. 
However, the related works on \gls{hevc} tile partitioning only address independently minimization of encoding time and encoding quality loss.
%, without trying to combine them as a trade-off. 
   
%In this paper, we propose a dynamic slice partitioning solution under \gls{vtm}-6.2 software.
In this work, we take advantage of the increased flexibility offered by the \glspl{rs} in \gls{vvc}, in order to propose a dynamic \gls{trs} partitioning solution under \gls{vtm}-6.2 software. 
Prior to the encoding of a frame, the \gls{trs} partitioning stage uses the spatial information and the times of previously encoded \glspl{ctu} in order to optimize the \gls{trs} partitioning.
%Prior to the encoding of every frame, the slice partitioning adjusts the slice partitioning to frame spatial information and to encoding times of previously encoded \glspl{ctu}.
The proposed solution minimizes a trade-off between encoding time and encoding video quality, which is a novel approach compared to related works.
%The proposed dynamic \gls{trs} partitioning stage establishes a trade-off between encoding time and encoding quality.
%This approach is novel compared to related works.
Moreover, to the best of our knowledge, this is the first work that implements a multi-thread \gls{vvc} reference encoder, generating baseline results for future related works.

The rest of the paper is organized as follows. Section~\ref{section:algo_description} describes the proposed solution, which establishes the trade-off between encoding time and encoding quality.
Section~\ref{section:Results} presents and analyses the experimental results on \gls{vtm}-6.2. Finally, Section~\ref{section:Conclusion} concludes this paper.

%%%%%%%%%%%%%%%%%%%%%%%%%%%%%%%%%%%%%%%%%%%%%%%%%%%%%%%%%%%%%%%%%%%%%%%%%%%%%%%

\section{Dynamic Frame Partitioning for Parallel Processing}\label{section:algo_description}

As mentioned in Section~\ref{section:intro}, the proposed \gls{trs} partitioning solution addresses simultaneously the minimization of encoding time and the limitation of encoding quality loss.
%During the slice partitioning search stage, that takes place before the encoding of every frame, a \g
This section first describes the encoding time minimization of the current frame, using times of previously encoded co-located \glspl{ctu}.
%This section first describes how to exploit the times of previously encoded \glspl{ctu}, in order to minimize the estimated encoding time of the current frame.
The second subsection introduces the clustering of spatial information into the \glspl{rs} to limit the encoding quality loss.
%The second subsection introduces the \gls{rs} clustering of \hl{frame spatial information: TROP DE MOTS D'AFFILEE}, whose aim is to limit the encoding quality loss.
The last subsection describes the proposed solution, that establishes a trade-off between encoding time and encoding quality.

\subsection{Encoding Time Minimization}\label{subsection:speedup}

%Let $P$ be a slice partitioning of current frame.
%Let current frame be partitioned $P$ be a partitioning into $S$ slices of current frame.
Let $P$ be the partitioning of current frame into $n$ \glspl{rs}: 
$P = \{s_0, ... , s_{n-1}\}$. %, with $s_j$ the \glspl{rs} composing $P$.
%In the following, $T(P)$ is the multi-thread encoding time of current frame, with $N$ threads working simultaneously.
In the following, $T(P)$ is the encoding time of current frame partitioned with $P$, and simultaneously processed 
by $N$ threads in parallel (each thread entirely dedicated to encode a single \gls{rs}).
%Each thread is entirely dedicated to a single slice, therefore $T(P)$ is equivalent to the time took by the slowest thread to encode    
In this case, $T(P)$ is equal to the time required  by the slowest thread to encode his \gls{rs}.
Eq.~\ref{eq:time_partition} formally establishes $T(P)$, with $T(c_i)$ the encoding time of \gls{ctu} $c_i$ and $T(s_j)$ the encoding time of the \gls{rs} $s_j$. 
\begin{equation} \label{eq:time_partition}
\begin{split}
T(s_j) &= \sum_{c_i \in s_j} T(c_i), \\
T(P) &= \max_{s_j \in P} (T(s_j)).
\end{split}
\end{equation}
%\green{The speed-up offered by $P$ is the ratio between the sequential encoding time of the non-partitioned frame and $T(P)$.
%Therefore, the problem of speed-up maximization is equivalent to the minimization of $T(P)$, with $P$ in the set of possible slice partitioning of the frame.} 
%\red{Utilisation de workload balancing ??}
Eq.~\ref{eq:time_partition} shows that $T(P)$ is directly determined by the \gls{ctu} encoding times $T(c_i)$.
However, during the \gls{trs} partitioning stage, these values are not available, since the \gls{trs} partitioning stage takes place before the encoding of current frame.
%In order to overcome this deficit of information, an array of estimators noted $\tilde{T}(c_i)$ is given as an input to the slice partitioning search stage.
In order to overcome this lack of information, the values $T(c_i)$ are replaced during the \gls{trs} partitioning stage by estimated values noted $\tilde{T}(c_i)$.
%In this work, the estimated times $\tilde{T}(c_i)$ are extracted from the \gls{ctu} \hl{pas clair} times of previously encoded frame.

%The minimization of $T(P)$ is converted into the minimization of an approximate $\tilde{T}(P)$, described by Eq.~\ref{eq:time_estimation}.
%\begin{equation}\label{eq:time_estimation}
%\begin{split}
%\tilde{T}(s_j) &= \sum_{c_i \in s_j} \tilde{T}(c_i)\\
%\tilde{T}(P) &= \max_{s_j \in P} (\tilde{T}(s_j))
%\end{split}
%\end{equation}
Several related works~\cite{storch_speedup-aware_2016,koziri_heuristics_2017} define $\tilde{T}(c_i)$ as the encoding time of the co-located \gls{ctu} (located at the same spatial coordinates) in the closest temporal frame previously encoded. 
This choice is motivated by the temporal continuity of the video sequences content.
%However, authors in~\red{[2 refs]} show that $T(c_i)$ is more correlated with the co-located \gls{ctu} of the frame with the closest \gls{qp} value, \green{compared to the co-located \gls{ctu} of the closest temporal frame}.
%This is caused by the Group of Pictures structure~\red{[ref gop RA]} in Inter coding configuration, that allows a distinct \gls{qp} value depending on the frame temporal layer.
In \gls{ra} configuration, authors in~\cite{chan_improve_2017} have shown that $T(c_i)$ is more correlated with the times of the co-located \gls{ctu} in co-\gls{tl} frame, compared to the co-located \gls{ctu} of the closest temporal frame.
%The co-\gls{tl} frame refers to the closest temporal frame previously encoded, belonging to same temporal layer.
The co-\gls{tl} frame refers to the previously encoded frame belonging to same temporal layer.
This is caused by the shared coding parameters of frames at similar temporal level in the group of pictures structure defined by the \gls{ctc}~\cite{boyce_jvet-j1010:_2018}.
% file \emph{encoder\_randomaccess\_vtm.cfg}\footnote{\url{https://vcgit.hhi.fraunhofer.de/jvet/VVCSoftware_VTM/blob/VTM-6.2rc1/cfg/encoder_randomaccess_vtm.cfg}}.
%This structure allows a distinct \gls{qp} value and prediction type depending on the frame temporal layer.
Following the results of~\cite{chan_improve_2017}, the selected estimator $\tilde{T}(c_i)$ is defined as the encoding time of the co-located \gls{ctu} in the co-\gls{tl} frame. 
%The minimization of $T(P)$ is converted into the minimization of an estimated $\tilde{T}(P)$, taking $\tilde{T}(c_i)$ values as an input.
The encoding time minimization technique consists in the search of a \gls{trs} partitioning $P$ that minimizes the estimated $\tilde{T}(P)$, computed with $\tilde{T}(c_i)$ values as an input.
%Ou juste dire maintenant le but va etre de chercher la decoupe qui minimise T chapeau, ce qui garantira que le travail entre les slices est bien reparti.
%La pertinence de la repartition de travail est directement lie a la precision de l'estimateur.}

%VERSION LONGUE, Resultats sur estiamteurs
%The following experiment proves this affirmation.
%Let us define the accuracy $\alpha$ of a time estimator $\tilde{T}$ for a given \gls{ctu} $c_i$ as: $\alpha(\tilde{T}(c_i)) = \frac{|T(c_i) - \tilde{T}(c_i)|}{T(c_i)}$.\\
%The average value for $\alpha$ has been computed across all the \glspl{ctu} of the 32 first frames of \gls{fhd} sequences \emph{BasketballDrive}, \emph{BQTerrace} and \emph{MarketPlace}. The results are shown in \Table{??} according to the time estimator and the frame temporal layer.
%\red{Mettre tableau + expliquer pourquoi si grosses differences de accuracy entre layer pour le co-QP.}\\
%\red{Mettre une figure de GOP.}

\subsection{Limitation of Encoding Quality Losses}\label{subsection:quality}

\begin{figure}[htbp]
	\centering
	\includegraphics[width=1.0\linewidth]{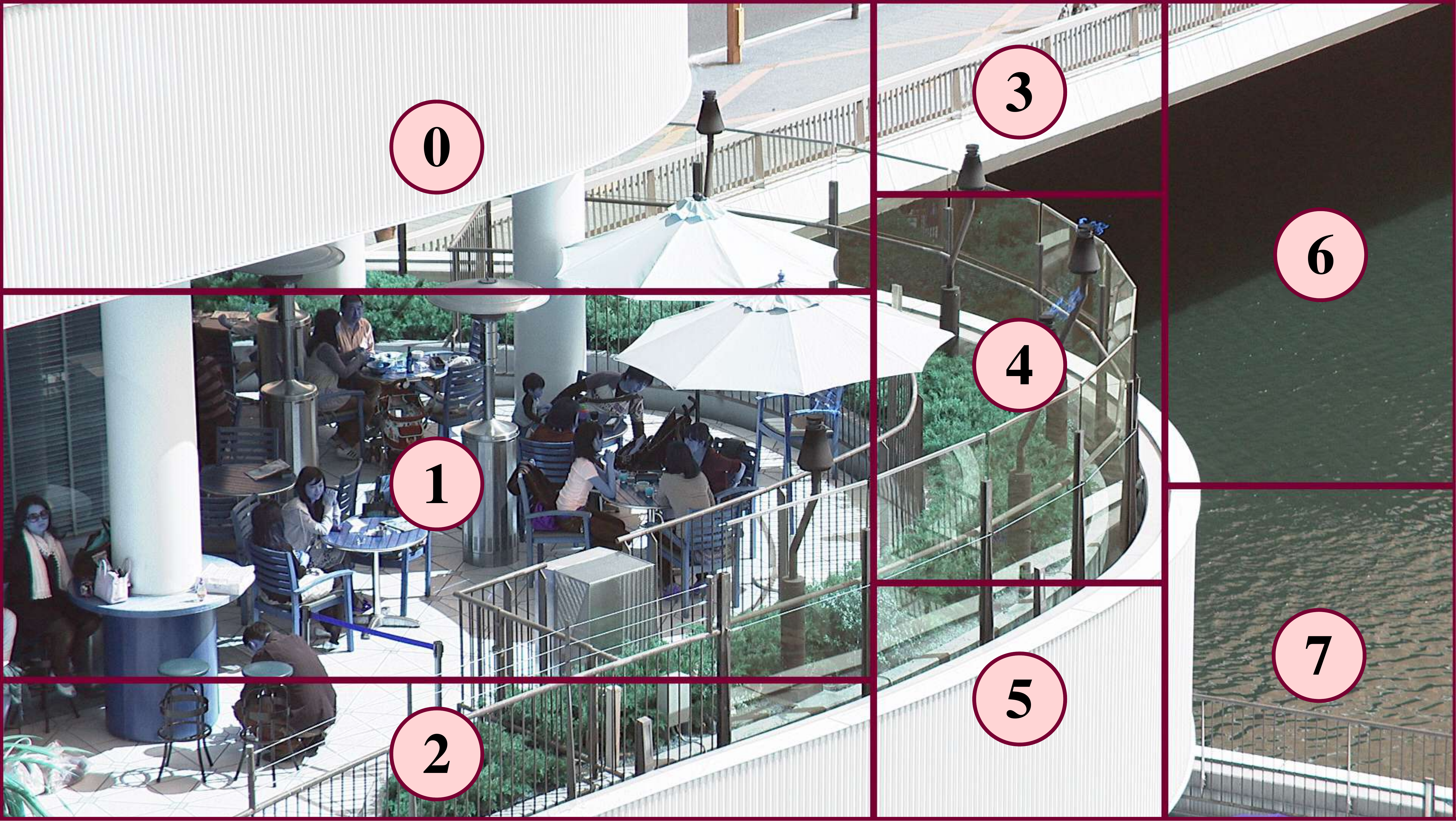}
	\caption{\gls{trs} partitioning of \emph{BQTerrace} frame \#4, computed with slice clustering.}
	\label{fig:Rect_clustering}
\end{figure}

%Algorithm AS 136: A k-means clustering algorithm
As mentioned in Section~\ref{section:intro}, prediction dependencies across \glspl{rs} boundaries are disabled and entropy coding state is reinitialized at each \gls{rs}.
In order to limit the encoding quality loss induced by these restrictions, the optimal \gls{trs} partitioning $P^*$ gathers similar spatial information inside the same \glspl{rs}.
%\red{(spatial information? pixel information ? luminance information ?)} 
This corresponds to a K-mean clustering~\cite{hartigan_algorithm_1979} of the spatial information into the \glspl{rs}, further called \gls{rs} clustering.
%Every cluster corresponds to a distinct \gls{rs}. 
The \gls{rs} clustering searches the \gls{trs} partitioning $P^*$ that minimizes the sum of luminance variance on all \glspl{rs}.
%Mathematically speaking, the \gls{rs} clustering searches the partitioning $P$ that minimizes the sum of \hl{luminance samples intra slice variances}.
%Mathematically speaking, the slice clustering minimizes the sum of luminance samples variances among the slices composing partitioning $P$.
Eq.~\ref{eq:quality_partition} computes the partitioning $P^*$ where $p_i$ is the value of luminance samples, and $\mu_j$ is the mean of \gls{rs} $s_j$ luminance samples.
%This partitioning $P$ is computed with Eq.~\ref{eq:quality_partition}, with $p_i$ is the value of luminance samples, and $\mu_j$ the mean of \gls{rs} $s_j$ luminance samples.
\begin{equation} \label{eq:quality_partition}
\begin{split}
P^* = \underset{P}{argmin} \left [\sum_{s_j \in P} \sum_{p_i \in s_j} \left (p_i -\mu_j \right )^2 \right ].
\end{split}
\end{equation}
%In other words, the K-mean clustering searches the slice partitioning $P$ that minimizes the sum of luminance samples variances among the slices composing to P.
%This technique obtains promising results in Chan et al work~\red{[ref]} for \gls{hevc} \red{(cheveu sur la soupe)}.

%\Figure{\ref{fig:Rect_clustering}} shows the 8 slices partitioning of \emph{BQTerrace} frame \#4, computed with slice clustering.
\Figure{\ref{fig:Rect_clustering}} shows the 8 \glspl{rs} partitioning, obtained by solving Eq.~\ref{eq:quality_partition} for frame \#4 of sequence \emph{BQTerrace}.
In \Figure{\ref{fig:Rect_clustering}}, regions of the frame with similar spatial information tend to be clustered into the same \glspl{rs}.
The dark water of the river is almost entirely contained in \glspl{rs} 6 and 7, and the light homogeneous regions of the frame are mainly included in \glspl{rs} 0, 3 and 5. 
On the other hand, the \glspl{rs} 1, 2 and 4 contain the regions with more complex spatial information.

\subsection{Two Steps Slice Partitioning Search}\label{subsection:tradeoff}

The \gls{trs} partitioning in \Figure{\ref{fig:Rect_clustering}} gathers similar spatial information inside the same \glspl{rs}, but is far from optimal regarding the encoding time minimization.
For instance, the encoding at $QP=27$ of \gls{rs}~\#1 is 12 times slower compared to the encoding of \gls{rs}~\#3, due among others to the greater area and spatial complexity of \gls{rs}~\#1 compared to \gls{rs}~\#3.
The encoding time of the considered frame is therefore sub-optimal due to the high encoding time of \gls{rs}~\#1. 
In order to reduce such imbalances between \glspl{rs} encoding times, the proposed solution combines the \gls{rs} clustering (Section~\ref{subsection:quality}) with the encoding time minimization technique (Section~\ref{subsection:speedup}).
%By combining the encoding time minimization technique with the \gls{rs} clustering, the proposed solution aims to reduce such imbalances between \glspl{rs} encoding times, without neglecting encoding quality.

\begin{figure}[htbp]
	\centering
	\includegraphics[width=1.0\linewidth]{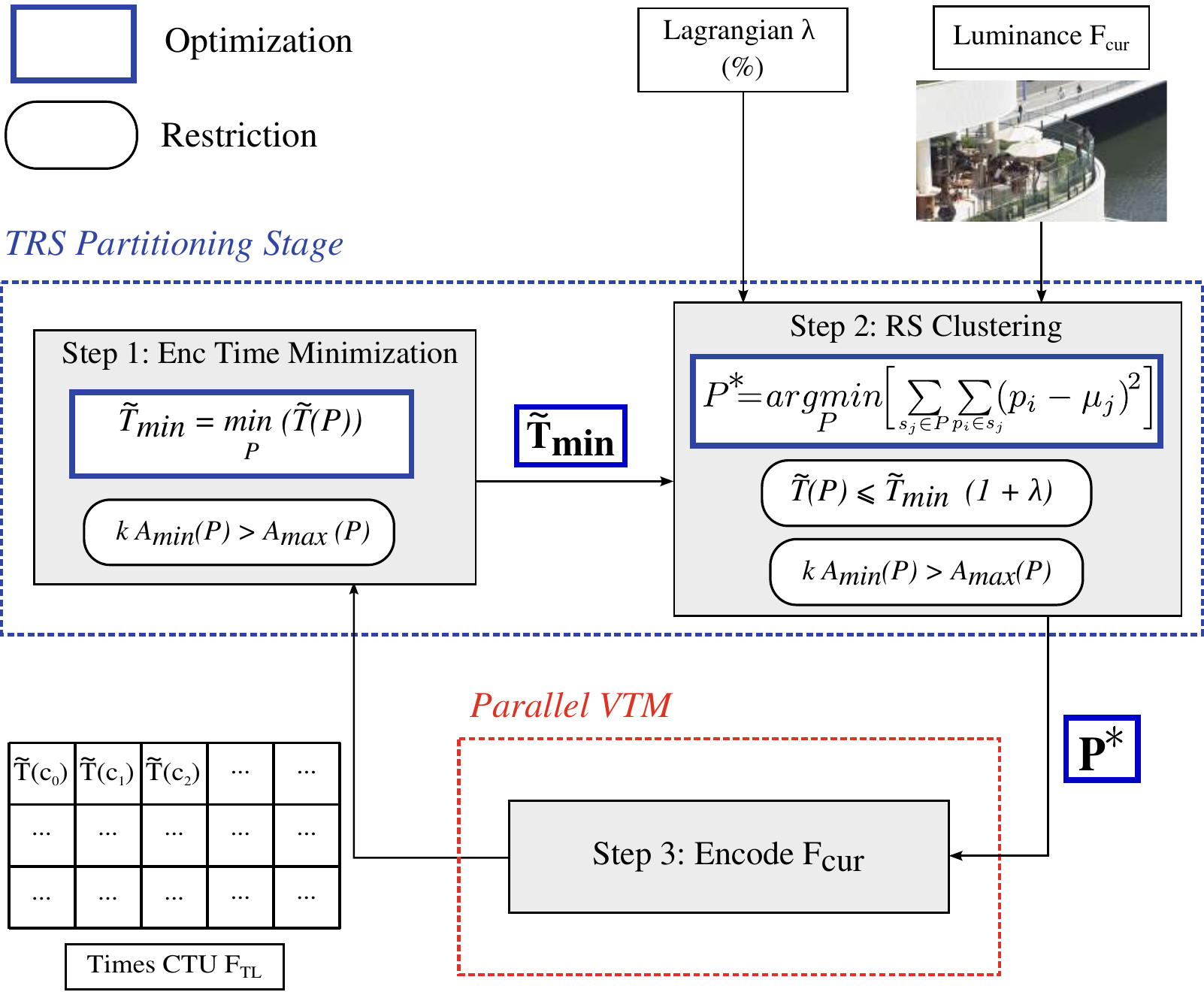}
	\caption{Proposed solution flowchart. }
	\label{fig:flowchart}
\end{figure}

%NOTES code : contraintes 
%verifier si numsplit coherent pour une tile 
%( airMeanCurBricks > 1.5 * airMeanRemainingBricks || airMeanCurBricks < airMeanRemainingBricks/1.5 )
% verifier si deux tailles de briques sont coherents, au niveau tile
% maxHeight > 2*minHeight
The proposed solution is represented as a flowchart in \Figure{\ref{fig:flowchart}}.
%Prior to the parallel encoding of current frame $F_{cur}$, the slice partitioning search stage is applied.
The \gls{trs} partitioning stage, enclosed in the blue dashed box, is applied prior to the parallel encoding of current frame $F_{cur}$, enclosed in the red dashed box. The \gls{trs} partitioning stage is divided into 2 distinct steps.  The first step is called encoding time minimization step.
This step computes the minimum estimated encoding time, defined by Equation~\ref{eq:est_time} and noted $\tilde{T}_{min}$.
\begin{equation} \label{eq:est_time}
\tilde{T}_{min} = \underset{P}{min}(\tilde{T}(P))
\end{equation}
The encoding time minimization step takes the \gls{ctu} times of the co-\gls{tl} frame $F_{TL}$ as input.\\
%\red{The \gls{ctu} times of the co-\gls{tl} frame $F_{TL}$ are needed as input.}\\

%The second step of the slice partitioning search computes the actual partitioning $P$, taking as input the minimum estimated time $\tilde{T}_{min}$, the luminance samples of $F_{cur}$, and a lagrangian parameter $t$ selected by the user that handles the trade-off between speed-up and encoding quality. 
%This second step computes the slice clustering of $F_{cur}$, under  
The second step of the \gls{trs} partitioning stage computes the \gls{rs} clustering of $F_{cur}$, under encoding time constraint. This step takes as inputs $\tilde{T}_{min}$ estimated during previous step, the luminance samples of $F_{cur}$, and a lagrangian parameter $\lambda$ that manages the trade-off between encoding time and encoding quality.
The possible values for $\tilde{T}(P)$ are bounded by Eq.~\ref{eq:lagrangian}.
\begin{equation}
\tilde{T}(P) \leq \tilde{T}_{min} \cdot (1+\lambda)
\label{eq:lagrangian}
\end{equation}
%The encoding time of the selected partitioning $P$ must not exceed: $ T(P) \leq T_{min} \cdot (1+\lambda)$.
When $\lambda=0$, only the partitioning $P$ that minimizes the estimated time is considered, since $\tilde{T}(P)=\tilde{T}_{min}$.
When $\lambda$ increases, more partitioning opportunities are offered to the \gls{rs} clustering, and therefore higher weight is given to encoding quality compared to encoding time minimization.
%When $t$ increases, the \gls{rs} clustering has a greater number of partitioning opportunities, and therefore more importance is given to encoding quality loss limitation.
The parameter $\lambda$ is therefore a means for the encoder to manage the trade-off, according to the requirement.

The aim of this paper is to show the relevance of a solution combining the 2 complementary steps previously presented.
For this reason, a near exhaustive search is conducted to compute both $\tilde{T}_{min}$ and \gls{rs} clustering.
As shown in \Figure{\ref{fig:flowchart}}, the only constraint given to the search algorithm: $k\cdot A_{min}(P) > A_{max}(P)$, with $A_{min}$ and $A_{max}$ the area of the smallest and the largest \glspl{rs}, respectively. 
The constant $k$ is set to $3$ in this work in order to contain search complexity.
The choice of less complex heuristics for the \gls{trs} partitioning stage is a distinct issue, that will be part of future works.
The global complexity overhead induced by the \gls{trs} partitioning stage is nonetheless measured and discussed further in this paper.
%An exhaustive search is conducted to compute both $\tilde{T}_{min}$ and \gls{rs} clustering.
%The only constraint in order to reduce search complexity is: $3\cdot A_{min}(P) > A_{max}(P)$, with $A_{min}$ and $A_{max}$ the area of the smallest and the largest \glspl{rs}, respectively. 
%Indeed, the aim of this paper is to show the relevance of a solution combining the 2 complementary steps previously presented.
%The choice of less complex heuristics for the \gls{trs} partitioning stage is a distinct issue, that will be part of future works.
%The global complexity overhead induced by the \gls{trs} partitioning stage is nonetheless measured and discussed further in this paper.

%%%%%%%%%%%%%%%%%%%%%%%%%%%%%%%%%%%%%%%%%%%%%%%%%%%%%%%%%%%%%%%%%%%%%%%%%%%%%%%
\section{Experimental Results} \label{section:Results}

This section presents the experimental setup, as well as the performance of the proposed \gls{trs} partitioning solution.

\subsection{Experimental Setup}

The following experiments are conducted under \gls{vtm}-6.2 software, built with gcc compiler version~7.4.0, under Linux version 4.15.0-74-generic  as distributed in Ubuntu-18.04.1.
The platform setup is composed of \glspl{cpu} Intel(R) Xeon(R) E5-2690 v3 clocked at 2.60~GHz, each of them disposing of 12 cores. The cores have each 768KB L1 cache, 3MB L2 cache and 30MB L3 cache.
%As mentioned in Section~\ref{subsection:speedup}, the number of \glspl{rs} is set to the number of threads.

The high-level parallelism structures included in \gls{vvc} standard allow to tackle complexity increase on multi-core processors. This complexity increase raises a critical issue mainly for high resolution video sequences.
For this reason, the test sequences selected in this work contain 4 \gls{uhd} and 5 \gls{fhd} sequences included in the \gls{ctc}~\cite{boyce_jvet-j1010:_2018}:
\emph{CatRobot1},
\emph{DaylightRoad2},
\emph{FoodMarket4},
\emph{Tango2} (\gls{uhd}), and
\emph{BQTerrace},
\emph{Cactus},
\emph{MarketPlace},
\emph{RitualDance} (\gls{fhd}).
The test sequences are encoded under \gls{ra} configuration at four \gls{qp} values: 22, 27, 32, 37.
The performance of our \gls{trs} partitioning solution is assessed by measuring the trade-off between the encoding quality using the \gls{bdr}~\cite{bjontegaard_calculation_2001} and the multi-thread speed-up $\sigma$, defined by Eq.~\ref{eq:speedup}.
\begin{equation}
\sigma = \dfrac{1}{4} \sum_{QP_i\in\{22,27,32,37\}}\dfrac{T_{O}(\gls{qp}_i)}{T_{R}(\gls{qp}_i)}
\label{eq:speedup}
\end{equation}

\noindent$T_{O}(\gls{qp}_i)$ and $T_{R}(\gls{qp}_i)$ are the original time (encoded with 1 \gls{rs} and 1 single thread) and reduced time (encoded with N \glspl{rs} and N threads) spent to encode the video sequence with $\gls{qp}_i$, respectively.
The overhead induced by \gls{trs} partitioning stage is further noted $\theta$ and measured in percentage of $T_{R}$.

\subsection{Performance of the Proposed Solution}

The theoretical upper bound in terms of speed-up, noted $\sigma_{max}$, for the proposed solution is computed with the Amdahl law~\cite{hill_amdahls_nodate}. Let $s$ be the sequential part (in $\%$) of an application.
The upper bound $\sigma_{max}$ obtainable with $n$ threads is expressed by Eq.~\ref{eq:amdhal}.
\begin{equation}
	\sigma_{max}(n) = \frac{1}{s+\frac{1-s}{n}}
	\label{eq:amdhal}
\end{equation}  
%In our case, the sequential portion of \gls{vtm}-6.2 encoder contains the stages handling data initialization, entropy, in-loop filter and bitstream writing.
In our case, the sequential portion of \gls{vtm}-6.2 encoder contains the data initialization, entropy, in-loop filter and bitstream writing stages.
All together, these stages represent $4\%$ of the encoding time in average across test sequences and \gls{qp} values.
Therefore, Eq.~\ref{eq:amdhal} provides the following upper bounds:
 $\pmb{\sigma_{max}(4)=3.57}$, $\pmb{\sigma_{max}(8)=6.25}$ and $\pmb{\sigma_{max}(12)=8.33}$.
%\begin{equation*}
% \sigma_{max}(4)=3.57, \sigma_{max}(8)=6.25, \sigma_{max}(12)=8.33.
%	\label{eq:amdhal}
%\end{equation*}  

As mentioned in Section~\ref{subsection:tradeoff}, the lagrangian parameter $\lambda$ manages the trade-off between encoding quality and encoding time minimization induced by the \gls{trs} partitioning.
Three values of parameter $\lambda$ (0, 0.1 and 0.3) are tested, and the one offering the best trade-off is selected according to thread number and resolution. 
\Table{\ref{tb:results_average}} presents the average results obtained with the selected $\lambda$ values, according to the resolution and number of threads $n$.
%\red{In the following, we present 3 distinct trade-offs, obtained with 3 values of $t$: 0\%, 10\% and 30\% (UN SEUL).}
Moreover, the results of the uniform \gls{trs} partitioning applied on the test sequences is also presented, in order to evaluate the performance of the proposed solution. The uniform \gls{trs} partitioning is an usual and straightforward technique that partitions the frame in a grid of the same \gls{rs} dimension. \\

\begin{table}[ht]
	\centering
	\caption{Average speed-up $\sigma$, \gls{bdr} and overhead $\theta$ obtained by both uniform and proposed \gls{trs} partitioning, according to the resolution and number of threads $n$.}
	\label{tb:results_average}
	\begin{adjustbox}{max width=1\columnwidth}
		%		\begin{adjustbox}{max width=1\textwidth}
		\begin{tabular}{@{}cc||c|c||c|c@{}}
			%				\toprule
			\multicolumn{2}{c}{}&\multicolumn{2}{c}{\textbf{FHD}} & \multicolumn{2}{c}{\textbf{UHD}}  \\\midrule
			&  &\textbf{Unif} &  \textbf{Proposed} & \textbf{Unif} &  \textbf{Proposed}  \\\midrule
			& 			&    	&  $\underline{\lambda = 0}$    &      & $\underline{\lambda = 0}$   \\
			&\gls{bdr} (\%)   			& 1.62	& \textbf{1.57}  & 1.31 	& \textbf{1.27}    \\
			$n=4$& Speed-up $\sigma$      & 2.68  & \textbf{3.10}  & 2.91 	& \textbf{3.27}   \\
			& $\theta (\%)$ 			&    	&  0.0  & 		& 0.0    \\\midrule
			
			&  	&   	& $\underline{\lambda = 0.3}$    &      & $\underline{\lambda = 0.1}$   \\
			&\gls{bdr} (\%)		& 2.69 	&  2.80	& 2.39	& \textbf{2.33}   \\
			$n=8$& Speed-up $\sigma$  	& 4.27  & \textbf{5.07}	& 4.55	& \textbf{5.34}   \\			
			& $\theta (\%)$ 	&  	& 0.01	&  	& 0.08		\\ \midrule
			
			&  &  	& $\underline{\lambda = 0.1}$    &      & $\underline{\lambda = 0.1}$   \\

			&\gls{bdr} (\%)	& 4.31	& \textbf{3.90} 	& 3.26	& \textbf{3.20}  \\
			$n=12$& Speed-up $\sigma$  & 5.57	& \textbf{6.44}  & 6.13	& \textbf{7.09}   \\
			& $\theta (\%)$ &  	& 0.54	&  	& 1.84		\\ \midrule
		\end{tabular}
	\end{adjustbox}
\end{table}

\Table{\ref{tb:results_average}} shows that the proposed \gls{trs} partitioning solution enables better results compared to uniform \gls{trs} partitioning in term of $\sigma$, regardless the resolution and number of threads $n$.
%The increase in $\sigma$ value ranges from $0.36$, for \gls{uhd} content with $n=4$, to $0.94$, for \gls{uhd} content with $n=12$.
The $\sigma$ increase ranges from $0.36$ to $0.94$, for \gls{uhd} content with $n=4$ and $n=12$, respectively.
The proposed \gls{trs} partitioning solution therefore reduces significantly the distance to the upper bounds $\sigma_{max}$ computed by Amdahl law, compared to uniform \gls{trs} partitioning.
This significant $\sigma$ increase proves the efficiency of the encoding time minimization step, presented in Section~\ref{subsection:speedup}.
It is important to note that the encoding time of every frame is reduced. Therefore both speed-up and latency are improved equally by the proposed solution.

In term of \gls{bdr}, the results of the proposed solution with the selected $\lambda$ values are slightly better (around $-0.05\%$) compared to uniform \gls{trs} partitioning.
Two exceptions are however noticeable. 
The \gls{bdr} decrease is substantial ($-0.41\%$) for \gls{fhd} content with $n=12$, and the only case for which the \gls{bdr} is slightly higher is for \gls{fhd} content with $n=8$ ($+0.11\%$). 
%In term of \gls{bdr}, the results are substantially better for \gls{fhd} content with $n=12$ ($-0.91\%$) compared to uniform \gls{trs} partitioning. 
%The only case for which the \gls{bdr} is slightly higher is for \gls{fhd} content with $n=8$ ($+0.11\%$).
%%The only case for which \gls{bdr} is slightly higher is for \gls{fhd} content with $n=8$ ($+0.11\%$).
%For all the the other cases, the \gls{bdr} decreases around $-0.05\%$ compared to uniform \gls{trs} partitioning.
The related works in \gls{hevc} minimizing the \gls{bdr} reported $0.16\%$~\cite{blumenberg_adaptive_2013} and $0.10\%$~\cite{chan_improve_2017} average \gls{bdr} decrease with 8 threads on \gls{fhd} and \gls{uhd} content.
%$-0.06\%$~\cite{jin_clustering-based_2016}
Our results in term of \gls{bdr} are therefore close to the results of previously mentioned works, even though these works minimize the \gls{bdr} without taking into consideration the speed-up optimization.
%The conclude can be used to limit the bdr increase, but not to achieve significant gains 
%

The conclusion of \Table{\ref{tb:results_average}} is that the proposed solution is able to maintain the \gls{bdr} increase to values close to uniform \gls{rs} partitioning. 
The variation of $\lambda$ value is however not sufficient to decrease significantly the \gls{bdr}, except for \gls{fhd} content with $n=12$. 
On the other hand, the proposed solution is highly effective to increase the speed-up offered by the \gls{trs} partitioning in \gls{vvc}.
% reducing significantly the distance to the upper bounds $\sigma_{max}$ provided by Amdahl law.
%The results presented in \Table{\ref{tb:results_average}} are indeed significantly closer to the upper bounds $\sigma_{max}$ provided previously in this section by Amdahl law.\\
Regarding the overhead $\theta$, the values are half induced by the encoding time minimization step, and half by the encoding quality loss limitation step.
The values are negligible when $n=4$ and $n=8$. 
For $n=12$, $\theta$ is greater than $0.5\%$ due to the almost exhaustive search implemented (see Section~\ref{subsection:tradeoff}). 
We are confident that the investigation of simple heuristics in future works will reduce greatly $\theta$, without degrading the results presented in \Table{\ref{tb:results_average}}.\\
%We are confident that $\theta$ will be greatly reduced by implementing less complex heuristics for the \gls{trs} partitioning stage.

\begin{table}[ht] %test_sequences
	\centering	
	\caption{Proposed solution with $\lambda = 0$ and $\lambda = 0.1$, encoded with 8 threads, according to \gls{uhd} sequence. }
	\label{tb:results_lambda0_lambda10}
	%	\begin{adjustbox}{max width=1\linewidth}
	\begin{adjustbox}{max width=1\columnwidth}
%		\begin{tabular}{@{}l||cc|cc|cc@{}}
		\begin{tabular}{@{}l||cc|cc@{}}			
			\multicolumn{5}{c}{\textbf{8 Threads, \gls{uhd} Sequences}} \\\midrule
			& 
%			\multicolumn{2}{c|}{\textbf{Uniform }} & 
			\multicolumn{2}{c|}{\textbf{Proposed Solution }} &
			\multicolumn{2}{c}{\textbf{Proposed Solution }}
			\\
			& 
%			\multicolumn{2}{c|}{} & 
			\multicolumn{2}{c|}{$\pmb{\lambda = 0 }$} &
			\multicolumn{2}{c}{$\pmb{\lambda = 0.1 }$}
			\\\midrule
			
			\multicolumn{1}{l||}{\textbf{Sequence}} & 
%			\begin{tabular}[c]{@{}c@{}}\gls{bdr} \\ (in \%) \end{tabular} &   \begin{tabular}[c]{@{}c@{}}\textbf{$\sigma$} \\  \end{tabular} & 
			
			\begin{tabular}[c]{@{}c@{}}\gls{bdr} \\ (in \%) \end{tabular} &   \begin{tabular}[c]{@{}c@{}}\textbf{$\sigma$} \\  \end{tabular} &   
			
			\begin{tabular}[c]{@{}c@{}}\gls{bdr} \\ (in \%) \end{tabular} &   \begin{tabular}[c]{@{}c@{}}\textbf{$\sigma$} \\  \end{tabular} 
				\\ \midrule
%\emph{CatRobot1},
%\emph{Tango2},
%\emph{FoodMarket4},
%\emph{DaylightRoad2} (\gls{uhd}), 
			\emph{CatRobot1}                                    &
			% 3840x2048                               &  
%			1.37 & 4.23 &
			 1.38& 5.24&
			 1.14& 5.19
			%			1.7 & 46.7 
			\\
			\emph{DaylightRoad}                 &
			% 3840x2048                               
%			1.84 & 4.92 &
			 1.82 &	5.79 &
			  1.70 & 5.70
			%			- & - 
			\\
			\emph{FoodMarket}                                  &
			% 3840x2048                               &  
%			3.79&4.04&
			4.09& 5.16&
			3.85& 5.10
			%			1.2 & 48.4 
			\\

			\emph{Tango2}                                     &
			% 2560x1600                               &  
%			2.55& 5.02&
			2.67& 5.54&
			2.61& 5.40
			%			- & - 
			\\\midrule

			\multicolumn{1}{l||}{\textbf{Average}} &
%			\textbf{2.39}  & \textbf{4.55} &
			\textbf{2.49}  & \textbf{5.43}  &
			\textbf{2.33} & \textbf{5.34}  \\\bottomrule
			%				\textbf{1.38}  &  \textbf{52.3}\\\bottomrule
		\end{tabular}
	\end{adjustbox}
\end{table}

\Table{\ref{tb:results_lambda0_lambda10}} shows the performance of the proposed solution with $\lambda = 0$ and $\lambda = 0.1$ running with 8 threads, according to the \gls{uhd} sequence.
%\blue{The notations $\Delta$\gls{bdr} and $\Delta \sigma$ represent our results relatively to the uniform \gls{trs} partitioning in term of \gls{bdr} and $\sigma$, respectively.}
As explained in Section~\ref{subsection:tradeoff}, the higher $\lambda$, the more importance is given to encoding quality with regard to the speed-up.
The results of \Table{\ref{tb:results_lambda0_lambda10}} are coherent with this explanation.
Indeed, for every sequence the proposed solution with $\lambda = 0.1$ enables better \gls{bdr} but lower $\sigma$ compared to the proposed solution with $\lambda = 0$.
%As mentioned in Section~\ref{subsection:tradeoff}, when $\lambda = 0\%$ only the encoding time minimization step is performed and the encoding quality loss is not considered.
%For every sequence, the proposed solution with $\lambda = 10\%$ enables better \gls{bdr} but lower speed-up $\sigma$ compared to the proposed solution with $\lambda = 0\%$.
%These results are coherent with Section~\ref{subsection:tradeoff}, where it is explained that the higher $\lambda$, the more importance is given to encoding quality with regard to the speed-up.
%Moreover, \Table{\ref{tb:results_lambda0_lambda10}} shows that the proposed solution with $\lambda=10\%$ reduces $\delta$\gls{bdr} to values close to 0, without degrading significantly the speed-up.
%This behavior is particularly noticeable for sequence \emph{FoodMarket}.
In average, the \gls{bdr} is 0.16\% better when selecting $\lambda = 0.1$, without degrading significantly $\sigma$ (-0.09).  
The results are particularly noticeable for sequence \emph{FoodMarket}.
For this sequence, the \gls{bdr} is 0.24\% better and $\sigma$ only decreases by 0.06\% when selecting $\lambda = 0.1$, compared to the proposed solution with $\lambda = 0$.
%Indeed, when $\lambda$ value is zero, the $\sigma$ increase represents $+1.12\%$ and the \gls{bdr} raises $+0.30\%$, compared to the uniform \gls{trs} partitioning.
%By setting $\lambda$ to 10\%, the speed-up $\sigma$ is maintained to $+1.06\%$, while the \gls{bdr} decreases to $+0.06\%$.

%\vspace{-1em}
%%%%%%%%%%%%%%%%%%%%%%%%%%%%%%%%%%%%%%%%%%%%%%%%%%%%%%%%%%%%%%%%%%%%%%%%%%%%%%%
\section{Conclusion} \label{section:Conclusion}
In this paper, a dynamic \gls{trs} partitioning is proposed for next generation video standard \gls{vvc}.
%The proposed solution addresses both encoding time minimization and limitation of encoding quality loss by combining them as a trade-off \hl{il faut unifier ces notations voires l etat de l art}.
The proposed solution combines two techniques to minimize multi-thread encoding time and encoding quality loss, respectively.
A lagrangian parameter $\lambda$ is applied, allowing to select a trade-off between encoding time and encoding quality.
The experiments show that the proposed solution decreases significantly multi-thread encoding time, with slightly better encoding quality, compared to uniform \gls{rs} partitioning.
Future works will focus among other points on the improvement of the \gls{ctu} time estimator, used in the encoding time minimization step.
%Instead of simply relying on the co-located \gls{ctu} times in the co-\gls{tl} frame, the motion information could provide \glspl{ctu} more similar to current \gls{ctu} compared to co-located \gls{ctu}.
Instead of simply relying on the co-located \gls{ctu} times of the co-\gls{tl} frame, future solutions will rely on \gls{ctu} deduced by motion information.
The investigation of lightweight heuristics for the \gls{trs} partitioning stage will also be part of future works. We are confident they will reduce drastically the overhead, especially for 12 threads encodings of \gls{uhd} content.

\def\url#1{}
\bibliographystyle{IEEEbib}
\bibliography{Bibliography}
\end{sloppypar}

\end{document}

%% file: Glossary.tex
\newacronym{avc}{AVC}{Advanced Video Coding}
\newacronym{ai}{AI}{All Intra}
\newacronym{alf}{ALF}{Adaptive Loop Filter}
\newacronym{aom}{AOM}{Alliance for Open Media}
\newacronym{amvp}{AMVP}{Advanced Motion Vector Prediction}

\newacronym{bt}{BT}{Binary Tree}
\newacronym{bh}{BTH}{Binary Tree Horizontal}
\newacronym{bv}{BTV}{Binary Tree Vertical}
\newacronym{bdr}{BD-BR}{Bj\o ntegaard Delta BitRate}
\newacronym{bdpsnr}{BD-PSNR}{Bj\o ntegaard Delta PSNR}
\newacronym{bdof}{BDOF}{Bi-Directional Optical Flow}

\newacronym{ctu}{CTU}{Coding Tree Unit}
\newacronym{cu}{CU}{Coding Unit}
\newacronym{cnn}{CNN}{Convolution Neural Network}
\newacronym{ctc}{CTC}{Common Test Conditions}
\newacronym{crc}{CRC}{Complexity Reduction Configuration}
\newacronym{cpu}{CPU}{Central Processing Unit}
\newacronym{cabac}{CABAC}{Context Adaptive Binary Arithmetic Coding}
\newacronym{CTC}{CTC}{Common Test Conditions}
\newacronym{ccitt}{CCITT}{International Telegraph and Telephone Consultative Committee}
\newacronym{cb}{CB}{Coding Block}

\newacronym{dl}{DL}{Deep Learning}
\newacronym{DCT}{DCT}{Discrete Cosine Transform}
\newacronym{DST}{DST}{Discrete Sine Transform}
\newacronym{dmvr}{DMVR}{Decoder-side Motion Vector Refinement}
\newacronym{dpb}{DPB}{Decoded Picture Buffer}
\newacronym{dbf}{DBF}{Deblocking Filter}

\newacronym{fhd}{FHD}{Full High Definition}
\newacronym{fps}{fps}{Frames Per Second}

\newacronym{gop}{GOP}{Group of Pictures}

\newacronym{hevc}{HEVC}{High Efficiency Video Coding}
\newacronym{hdr}{HDR}{High Dynamic Range}
\newacronym{hi}{HI}{Horizontal Inconsistency}
\newacronym{isp}{ISP}{Intra Sub-Partitionning}
\newacronym{hm}{HM}{HEVC Model}
\newacronym{hd}{HD}{High Definition}
\newacronym{HFR}{HFR}{High Frame Rate}
\newacronym{hvs}{HVS}{Human Visual System}

\newacronym{iso}{ISO}{International Organization for Standardization}
\newacronym{iec}{IEC}{International Electrotechnical Commission}
\newacronym{itu}{ITU}{International Telecommunication Union}

\newacronym{jvet}{JVET}{Joint Video Exploration Team}
\newacronym{jem}{JEM}{Joint Exploration Model}

\newacronym{qt}{QT}{Quad Tree}
\newacronym{qp}{QP}{Quantization Parameter}
\newacronym{qtbt}{QTBT}{Quad Tree Binary Tree}

\newacronym{lmcs}{LMCS}{Luma Mapping with Chroma Scaling}
\newacronym{ld}{LD}{Low-Delay}
\newacronym{ldp}{LDP}{Low-Delay P}
\newacronym{ldb}{LDB}{Low-Delay B}

\newacronym{mpm}{MPM}{Most Probable Mode}
\newacronym{mv}{MV}{Movement Vector}
\newacronym{ml}{ML}{Machine Learning}
\newacronym{mi}{MI}{Mutual Information}
\newacronym{mdf}{MDF}{Motion Divergence Field}
\newacronym{mc}{MC}{Motion Compensation}
\newacronym{me}{ME}{Motion Estimation}
\newacronym{mpeg}{MPEG}{Moving Picture Experts Group}
\newacronym{mtt}{MTT}{Multi-Type Tree}
\newacronym{MOS}{MOS}{Mean Opinion Score}
\newacronym{MTS}{MTS}{Multiple Transform Selection}
\newacronym{mip}{MIP}{Matrix-based Intra Prediction}
\newacronym{mimd}{MIMD}{Multiple Instructions on Multiple Data}
\newacronym{mse}{MSE}{Mean Squared Error}

\newacronym{pdf}{Pdf}{Probability Density Function}
\newacronym{pu}{PU}{Prediction Unit}
\newacronym{psnr}{PSNR}{Peak Signal to Noise Ratio}
\newacronym{pmd}{PMD}{Pyramid Motion Divergence}
\newacronym{poc}{POC}{Picture Order Count}
\newacronym{pps}{PPS}{Picture Parameter Set}

\newacronym{rd}{RD}{Rate Distorsion}
\newacronym{rdo}{RDO}{Rate Distorsion Optimization}
\newacronym{ra}{RA}{Random Access}
\newacronym{rgb}{RGB}{Red Blue Green}
\newacronym{rf}{RF}{Random Forest}
\newacronym{rs}{RS}{Rectangular Slice}

\newacronym{satd}{SATD}{Sum of Absolute Transform Differences}
\newacronym{svm}{SVM}{Support Vector Machines}
\newacronym{si}{SI}{Spatial Information}
\newacronym{scc}{SCC}{Screen Coding Content}
\newacronym{simd}{SIMD}{Single Instruction on Multiple Data}
\newacronym{shvc}{SHVC}{Scalable High Efficiency Video Coding}
\newacronym{SAO}{SAO}{Sample Adaptive Offset}
\newacronym{SPS}{SPS}{Sequence Parameter Set}
\newacronym{sisd}{SISD}{Single Instruction on Single Data}
\newacronym{ssim}{SSIM}{Structural SIMilarity}
\newacronym{sbtmvp}{SbTMVP}{Subblock-based Temporal Motion Vector Prediction}
\newacronym{samviq}{SAMVIQ}{Subjective Assessment Methodology for Video Quality}

\newacronym{tu}{TU}{Transform Unit}
\newacronym{ti}{TI}{Temporal Information}
\newacronym{tt}{TT}{Ternary Tree}
\newacronym{tth}{TTH}{Ternary Tree Horizontal}
\newacronym{ttv}{TTV}{Ternary Tree Vertical}
\newacronym{trs}{TRS}{Tile and Rectangular Slice}
\newacronym{tl}{TL}{Temporal Layer}

\newacronym{uhd}{UHD}{Ultra High Definition}

\newacronym{vvc}{VVC}{Versatile Video Coding}
\newacronym{vi}{VI}{Vertical Inconsistency}
\newacronym{vceg}{VCEG}{Video Coding Experts Group}
\newacronym{vtm}{VTM}{VVC Test Model}
\newacronym{vmaf}{VMAF}{Video Multi-Method Assessment Fusion}

\newacronym{wpp}{WPP}{Wavefront Parallel Processing}
\newacronym{waip}{WAIP}{Wide Angular Intra Prediction}